\documentclass{tADP2e}

\usepackage{epstopdf}
\usepackage{lmodern}

\usepackage{color}

\newcommand{\rem}[1]{}
\newcommand{\figwid}{0.48\textwidth}

\begin{document}

\title{Current trends in the physics of nanoscale friction}

\author{
  \name{
    N. Manini,\textsuperscript{a}
    G. Mistura,\textsuperscript{b}
    G. Paolicelli,\textsuperscript{c}
    E. Tosatti,\textsuperscript{d,e,f}
    and
    A. Vanossi\textsuperscript{e,d} \thanks{vanossi@sissa.it}
  }
  \affil{
    \textsuperscript{a}Dipartimento di Fisica, Universit\`a degli Studi di
    Milano, Via Celoria 16, 20133 Milano, Italy
  }
  \affil{
    \textsuperscript{b}Dipartimento di Fisica e Astronomia ``G. Galilei'',
    Via Marzolo, 8 35131 Padova
  }
  \affil{
    \textsuperscript{c}CNR-Istituto Nanoscienze - Centro S3,
    Via Campi 213 41100, Modena, Italy.
  }
  \affil{
    \textsuperscript{d}International School for Advanced Studies (SISSA),
    Via Bonomea 265, 34136 Trieste, Italy
  }
  \affil{
    \textsuperscript{e}CNR-IOM Democritos National Simulation Center,
    Via Bonomea 265, 34136 Trieste, Italy
  }
  \affil{ 
    \textsuperscript{f}International Center for Theoretical Physics (ICTP),
    Strada Costiera 11, 34151 Trieste, Italy
  }
}

\maketitle

\begin{abstract}
Tribology, which studies surfaces in contact and relative motion, includes friction, 
wear, and lubrication, straddling across different fields: mechanical engineering, 
materials science, chemistry, nanoscience, physics. 
This short review restricts to the last two disciplines, with a qualitative survey 
of a small number of recent progress areas in the physics of nanofriction.
\end{abstract}


\section{Introduction}
\label{introduction:sec}

From the elemental surface sliding of an atomically sharp tip, to the
squeaking of door hinges, all the way up to the complex evolution of a
geophysical fault, friction abounds in nature -- spanning, in disparate
areas, vastly different scales of length, time, and energy.
Despite the fundamental, practical and technological importance of
tribology, several key physical aspects of mechanical dissipative phenomena
are not yet fully understood, mostly due to the complexity of highly
out-of-equilibrium nonlinear processes often occurring across
ill-characterized sliding interfaces. For centuries, and until quite
recently, scientists made only modest inroad on the atomic-level physics
involved in frictional processes, leaving developments largely to empirism
and engineering.
With the ongoing quest for ``holy grails'' such as the control of friction
by atomistic design, or the hopeful gap-bridging across the different
scales, reaching a macroscopic description as it may emerge from the
fundamental atomic principles, new avenues of research are being pursued
and new discoveries are being made -- and that, especially at the nano/meso
scales, thanks to remarkable developments in nanotechnology.
Progress at the fundamental physics level is going on both through
nanofriction experiments, and through theory from computer simulations to
non-equilibrium statistical mechanics.
Far from covering any of that exhaustively, we merely intend to provide a
glance at some themes which in our view contain the seed for further work,
and whose development is familiar through our recent research involvement.

In that spirit we will therefore cover:  the elementary processes of dry,
wearless surface sliding (Sec.~\ref{depinning:sec});
structural lubricity -- sometimes called ``superlubricity''
(Sec.~\ref{superlub:sec}); thermolubricity (Sec.~\ref{thermo:sec});
tribological properties of layered, graphitic-like, materials
(Sec.~\ref{layered:sec}); electronic, magnetic,
and quantum effects in friction (Sec.~\ref{exotic:sec});
artificial frictional systems with colloids and ions in optical lattices
(Sec.~\ref{optical:sec}).

\section{Static and Kinetic Friction: from Depinning to Sliding}
\label{depinning:sec}

A slider at rest generally requires a finite force, the static friction
force in order to start sliding.
Subsequently, a different force, the kinetic or dynamic friction force,
often smaller than the static one, needs to be applied in order to maintain
a steady sliding motion.

At face value, the transition from a static strained configuration to full
sliding is conceptually as simple as overcoming an energy barrier. However,
practical single- and multiple-contact conditions are characterized by
complex interaction profiles plus nontrivial internal dynamics.
As a result, the interplay of thermal drifts, contact ageing,
contact-contact interactions, and macroscopic elastic deformations
introduce significant complications, and make the depinning transition from
static to kinetic friction an active field of research. The depinning
dynamics affects in particular the transition between stick-slip and smooth
sliding for sliding friction.
Following and extending previous literature, F.P.\ Landes {\it et al.}
\cite{Landes15} address this problem in the context of the viscoelastic
dynamics of a spring-block system driven across a rough surface.
The proposed mean-field model provides a microscopic basis for the
macroscopic description in terms of rate-and-state equations.  In several
works by J.\ Fineberg's group \cite{Rubinstein04, Rubinstein06, Rubinstein12}
the transition from sticking to sliding is characterized by slip fronts
propagating along the interface. Several works have modeled this
transition using various techniques including a master-equation type of
approach \cite{BP2011, BP2012, Braun12a, BP2013, BT2014, Braun15},
mesoscopic models \cite{Tromborg11}, and finite-elements techniques
\cite{Kammer12, Taloni15, Kammer16}.

The same fundamental problem is being investigated in the framework of
lubrication. Mutually sliding macroscopic machines parts need to be kept
lubricated, typically by mineral-based oils to maintain normal
operation, and the same is required for the systems at a microscopic
scale. An especially efficient state of lubrication for macroscopic
machines is that of hydrodynamic lubrication, with surfaces separated by
a relatively thick liquid lubricant film so that direct surface contact
and wear is prevented. As a result the friction coefficient due to the
shear of liquid films could be relatively low ($\sim 0.005$). In practice,
machines operate mostly in a state of boundary lubrication, where the
lubricating film is mostly squeezed out, and the surfaces are protected
by films of adsorbed lubricant molecules, often sticking there thanks to
chemical reactions.

Down to the microscopic or nanometer scale, hydrodynamic lubrication is
difficult to achieve because lubricants are unlikely to flow through a
nanometer gap because of the increased effective viscosity and a
tendency to layering or even solidification \cite{Persson94a, Gao97a, Persson00}. 
Traditional mineral oils with additives used for boundary
lubrication are likely to fail in microscopic contacts for two reasons:
(i) mineral oil has poorer lubrication properties for the silicon-based
materials widely used to build MEMS than for steel; (ii) for devices at
the nanometer scale, lubricant additives are easily of the same size as
that of the contact itself, so that the molecules could act rather as an
obstructor to the motion of nano-component than as a lubricant. As a
result, solid lubrication becomes an option for protecting the surfaces
in microscopic or nanometer-scale systems.

\begin{figure}
  \centerline{\includegraphics[width=\figwid]{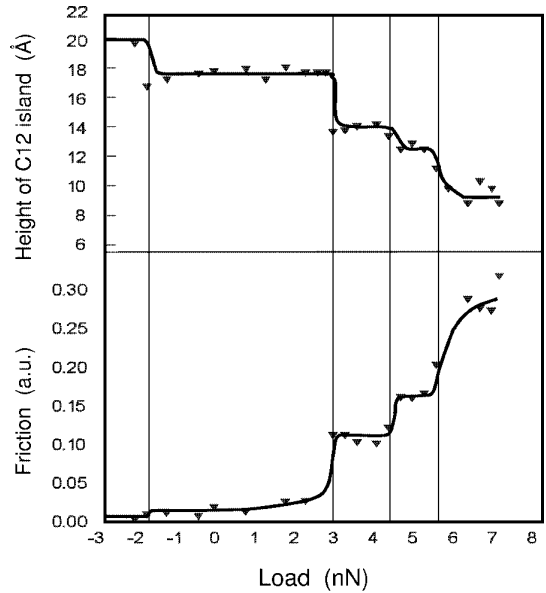}}
  \caption{\label{forcesteps:fig}
    Height (top) and friction force (bottom) measured on islands
    of C16 thiols on Au(111) as a function of applied load. Stepwise changes
    are observed at critical loads. Some heights in particular are more
    stable than others, as indicated by the length of the plateaus. From
    Ref. \cite{Salmeron01}, copyright 2001, Springer.
  }
\end{figure}

Ordered Langmuir-Blodgett (LB) films and self-assembled monolayers (SAMs)
are currently a popular choice for solid lubrication \cite{Liu16}. With
such kinds of lubricant, the dependence of friction on applied load and
sliding velocity was investigated in detail \cite{Zhang02}. Friction on
molecular films is usually significantly reduced compared to the bare
surface of the substrates, with typical friction coefficients ranging from
$0.05$ to $0.1$. Friction forces increase with increasing normal load, but
generally more slowly than predicted by Amonton's law. Interestingly, under an
increasing applied load, it is observed that the film molecules tilts only
at discrete angles,
due to the ratchet-like intermolecular binding and the zig-zaging C-C
skeleton in the film chains \cite{Salmeron01}. These discrete changes in
tilt angles result in a stepwise decrease in the film height and
corresponding increase in friction as illustrated in
Fig.~\ref{forcesteps:fig}.
Similar quantized changes in film thickness and friction forces are
observed also in confined liquid films \cite{Atkin07, Bovio09a, Perkin14},
and should therefore be considered as a fundamental feature of
highly-ordered closely-packed molecular films \cite{Capozza15a}.

In the presence of surface films, the dependence of friction on the
sliding velocity is more intricate. Low-speed stick-slip dynamics of
SAM-covered mica surfaces would evolve into smooth sliding when the
sliding velocity exceeds a certain threshold \cite{Berman96}.
In Atomic Force Microscopy (AFM) experiments, when the tip scans over the
monolayers at low speeds, friction force is reported to increase with the
logarithm of the velocity, similar to that observed when the tip scans
across crystalline surfaces. This velocity dependence is interpreted in
terms of thermally activated depinning of interlocking barriers involving
interfacial atoms \cite{Gnecco00}.

\section{Contact Area Dependence and New Perspectives in Superlubricity}
\label{superlub:sec}

The dependence of friction on the contact area stands at the heart of the
quantitative understanding of tribology, and it is interlinked to the
theory of the load dependence.
The macroscopic Da Vinci-Amontons law -- friction independent of area -- is
not confirmed at the microscopic scale.
In most nanoscale investigations the friction of a single contact is found
to increase linearly with the contact area \cite{Sheehan96, Ritter05,
  Dietzel08}.
In contrast, structurally mismatched atomically flat and hard crystalline
or amorphous surfaces are expected to produce a sublinear increase of
friction with contact area.
The frequent finding of friction proportional to area even in some of these
cases can be understood as a consequence of softness, either if the
interface, or of surface contaminants leading to effectively
pseudo-commensurate interfaces \cite{Muser01, He99}.
A systematic investigation of the dependence of friction on the contact
area was carried out recently for nanosized metal clusters on graphite in
ultraclean and even atmospheric conditions \cite{Dietzel13, Cihan16}, whose
results indicate a size dependence of kinetic friction which is scattered.
For certain clusters a regular linear scaling with area is observed, while
others can be grouped in sets compatible with sublinear scaling, which
match the expectations for structural lubricity \cite{Kim09, ManiniBraun11,
  deWijn12, Braun13, Koren16a, Koren16b}, sometimes also called
superlubricity.

Superlubricity, now a pervasive concept of modern tribology, dates back to
the mathematical framework of the Frenkel Kontorova model for
incommensurate interfaces \cite{Peyrard83}. When two contacting crystalline
workpieces are out of registry, by lattice mismatch or angular
misalignment, the minimal force required to achieve sliding, i.e. the
static friction, tends to zero in the thermodynamic limit -- that is, it
can at most grow as a power less than one of the area -- provided the two
substrates are stiff enough.
A parallel reasoning may apply to hard amorphous interfaces
\cite{Muser01,Monti17}.
%
These geometrical configurations prevents asperity interlocking and
collective stick-slip motion of the interface atoms, with a consequent
negligibly small frictional force. Practically, systems achieving low
values of dry sliding friction are of great technological interest to
significantly reduce dissipation and wear in mechanical devices functioning
at various scales.

\begin{figure}
  \centerline{\includegraphics[width=\figwid]{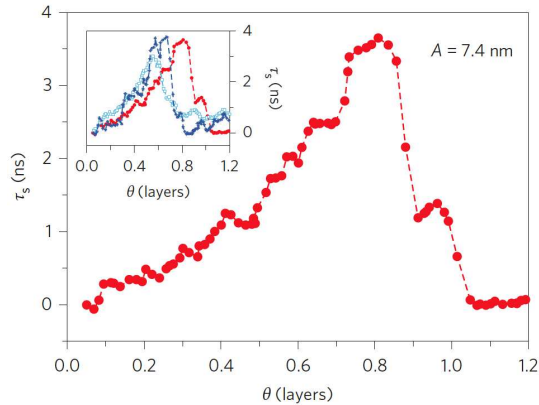}}
  \caption{\label{QCM:fig}
    Slip time of Xe on Cu(111) as a function of film coverage. The scan is
    taken at $T=47$~K with a quartz crystal microbalance oscillating at
    5~MHz with an amplitude of the Cu electrode of 7.4~nm. The coverage is
    deduced from the frequency shift, assuming for the monolayer an areal
    density corresponding to the completion of the $\sqrt{3}\times \sqrt{3}$
    commensurate solid phase.
    The data show remarkably large slip times with increasing submonolayer
    coverage, which are attributed to superlubricity of the incommensurate
    Xe islands, followed by a dramatic drop to zero for the dense
    commensurate monolayer. Inset: scans of Xe on Cu(111) taken for
    different Xe depositions on the same substrate at the same amplitude
    and temperatures between 47 and 49~K.
    The observed erratic behavior is associated to the first-order nature
    of the 2D density jump which destroys superlubricity with increasing
    adsorbate coverage near one monolayer.
    As such, it is expected to occur with hysteresis, which implies a
    difference between atom addition and atom removal, as well as
    occasional differences between one compressional event and another.
    From Ref. \cite{Pierno15}, copyright 2015, Macmillan Publishers Limited.
}
\end{figure}

Superlubricity is experimentally rare. Until recently, it has been
demonstrated or implied in a relatively small number of cases
\cite{Cieplak94, Dienwiebel04, Filippov08, Dietzel08, Schirmeisen09,
  Brndiar11}.
There are now more evidences of superlubric behavior in cluster
nanomanipulation \cite{Dietzel13, Guerra16, Cihan16}, sliding colloidal
layers \cite{Bohlein12, Vanossi12, Vanossi12PNAS}, and inertially driven
rare-gas adsorbates \cite{Pierno15, Varini15} (see Fig.~\ref{QCM:fig}).

Moreover there exists a vast literature discussing superlubric sliding
effects in the context of graphite/graphene flakes on a graphitic
substrate.
Free movements of graphene nanoflakes on graphene are observed at low
temperature and in UHV conditions using Scanning Tunneling Microscopy (STM)
\cite{Feng13}. This system exhibits a completely commensurate or
incommensurate interface as a function of the misfit angle and a systematic
computational study of the interlayer interaction energy in a rigid
bi-layer graphene system \cite{Xu13} confirms that interaction energy is
nearly constant for all incommensurate configurations.
By taking into account the flexibility of the graphene nanoflakes, and
their tribological response under the action of increasing normal loads,
atomistic simulations \cite{Bonelli09,vanWijk13} show that the smooth
sliding dynamics of these type of adsorbates may turn into a more
dissipative stick-slip regime of motion.

A breakdown of structural lubricity may occur at the heterogeneous
interface of graphene and h-BN. Because of lattice mismatch (1.8\%),
this interface is intrinsically incommensurate, and superlubricity should
persist regardless of the flake-substrate orientation, and become more 
and more evident as the flake size increases \cite{Leven13}.
However, vertical corrugations and planar strains may occur at the
interface even in the presence of weak van der Waals interactions and,
since the lattice mismatch is small, the system can develop locally
commensurate and incommensurate domains as a function of the misfit angle
\cite{Woods14, Guerra17}.
%
Nonetheless, spontaneous rotation of large graphene flakes on h-BN is
observed after thermal annealing at elevated temperatures, indicative of
very low friction due to incommensurate sliding \cite{Woods16, Wang16}.

Structural superlubricity over micrometer length scale is reported on
HOPG graphite \cite{Liu12,  Jiarui13}.
In these experiments, the cap of micron-diameter graphite pillars is
dragged laterally, producing a shear movement of the upper part relative to
the base. After releasing the cap, a self-retraction movement is
systematically found that leads the sheared mesas back to the original
position. A similar result but at smaller length scales \cite{Koren15} is
obtained by laterally moving the tip of an AFM microscope glued to a HOPG
graphite cylindrical pillar to cut and shear a planar section of the pillar
with respect to its base (cylinder radii between 50 and 300~nm).
This system develops a shear force which is composed by two parts, a
reversible component connected to the relative displacement and a smaller
irreversible part identified as the frictional response to sliding, which
is attributed to stochastic events due to the interaction of incommensurate
interface lattices.

\begin{figure}
  \centerline{\includegraphics[width=\figwid]{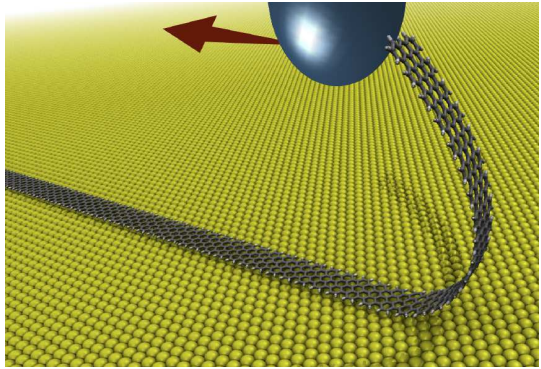}}
  \caption{\label{GNR:fig}
    Atomistic simulation of a graphene nanoribbon (GNR) deposited over a
    gold (111) surface and driven by an AFM tip, resembling the
    experimental setup of Ref.~\cite{Kawai16}. Depending on the effective
    mechanical instability induced by the lifted part, both peeling and
    sliding of the GNR are possible.
  }
\end{figure}

The current capability of synthesizing and manipulating quasi-1D atomically
perfect objects of extended length, such as telescopic nanotubes
\cite{Zhang13, Nigues14, Zhang16}, graphene nanoribbons \cite{Cai10,
  Ruffieux16, Kawai16}, aromatic polymers \cite{Kawai14}, or soft
biological filaments \cite{Ward15}, open now the possibility to transpose
the peculiar nanoscale tribological properties to extended contacts and
exploit them to control sliding-induced energy dissipation in
state-of-the-art technological devices (see Fig.~\ref{GNR:fig}).

Other experiments \cite{Berman15} have achieved vanishingly small friction
coefficients in dry macroscale sliding contacts by adding graphene in
combination with crystalline diamond nanoparticles. Simulations show
that wrapping of graphene patches around the tiny nanodiamonds lead to
nanoscrolls with reduced contact area that slide easily against an
amorphous diamondlike carbon surface.

In such expanding scenario toward larger scales, the robustness of the
superlubricity phenomenon remains a challenge, and the conditions of its
persistence or the mechanisms leading to its failure are cast as key
questions to be addressed.
Indeed, we know from theory and simulation \cite{Muser04, Ma15, Sharp16}
that even in clean wearless friction experiments with perfect atomic
structures, superlubricity at large scales may, for example, surrender due
to the soft elastic strain deformations of contacting systems.

\section{Temperature Dependence and Thermolubricity}
\label{thermo:sec}

In classical physics, all barriers are surmounted by thermal fluctuations
at finite temperature, provided one could wait long enough. While this
statement may easily be academic for macroscopic sliders, where the waiting
time could exceed the age of the universe, the consequence for microscopic
or nanoscopic sliders is that friction must vanish in the limit of zero
sliding velocity, then growing linearly at nonzero velocity.
This is the essence of the substantial suppression of friction due to
thermal effects, referred to in recent times as thermolubricity
\cite{Krylov05}.
L. Prandtl first recognized the role of temperature in reducing friction
\cite{Prandtl28}. In general, the energy landscape of an AFM tip dragged
along an atomically flat substrate exhibits valleys and barriers, and
thermal excitations at sufficiently low speed always provide sufficient
energy to overcome local barriers and enable slip \cite{Krylov14}.
Thus, it is commonly expected that the friction of a dry nanocontact should
classically decrease with increasing temperature provided no other surface
or material parameters are altered by the temperature changes
\cite{Krylov05, Jinesh08, Szlufarska08, Steiner09, Jansen10}.

A breakdown of this simple rule is provided by Friction Force Microscopy
(FFM) experiments that find a peak in the wearless friction of a point
contact at cryogenic temperatures for several classes of materials,
including amorphous, crystalline, and layered surfaces \cite{Schirmeisen06,
  Barel10a}.
Simulations performed within the Prandtl-Tomlinson (PT) model reveal that
temperature can affect the slip length resulting in a nonmonotonic
temperature dependence of friction \cite{Tshiprut09}. Simulations best
representing the experimental conditions show that this dependence
emerges from two competing processes acting at the interface: the
thermally activated formation and the rupturing of an ensemble of atomic
contacts \cite{Barel10a,Barel10b}.
In addition, a new competing mechanisms due to athermal instability
inherent in AFM measurement has been proposed \cite{Dong12}. Simulations
taking into account only this effect show a friction plateau at cryogenic
temperatures.
In that limit of course one should also worry about quantum effects, an aspect
which has not received much attention so far. 

An extension of the PT model, incorporating the possibility of thermally
activated contact strengthening, reproduces the normal PT-like behavior
of the friction force at sufficiently low and high temperatures. In the
intermediate temperature range, the model explains the experimentally
observed possibility of a minimum and a maximum in the temperature
dependence \cite{Evstigneev13}.

Thermal effects have also been investigated using a Quartz Crystal
Microbalance (QCM).
A recent nanofriction experiment reveals that Xe monolayers are fully
pinned to a graphene surface at low temperature \cite{Pierno14}. Above
30~K, the Xe film slides. The depinning onset coverage beyond which the
film starts sliding decreases with temperature. Similar measurements
repeated on bare gold show an enhanced slippage of the Xe films and a
decrease of the depinning temperature below 25~K.
Molecular-dynamics simulations relying on ab-initio derived potentials
indicate that the key mechanism to interpret this thermolubric effect is
the size dependence of the island commensurability. The latter quantity is
deeply affected also by the lattice misfit, which explains the different
frictional behavior of Xe on graphene and gold \cite{Restuccia16}.
All aforementioned investigations are carried out under UHV condition,
where probe and sample can be prevented from being polluted. However, the
contacting surfaces of moving elements in technology are usually in
atmospheric environment and may be covered by very thin water
layers.
Meniscus bridges might then form when the neighboring surfaces are in
contact or close to each other. FFM experiments with cantilever probes
featuring an in situ solid-state heater report an increase in friction by a
factor of 4 in humid air varying the tip temperature from $25^\circ$C to
$120^\circ$C, while in dry nitrogen friction decreases by $\sim40$\%
\cite{Greiner10}.
These trends are attributed to thermally assisted formation of capillary
bridges between the tip and substrate in air, and thermally assisted
sliding in dry nitrogen \cite{Greiner10, Barel12}. Increasing the
temperature above $\sim 150^\circ$C drives water away from the
contact. The friction force then decreases substantially and it becomes
equal to that measured in dry nitrogen \cite{Greiner12}. The thermally
activated growth process of a capillary meniscus also affects the adhesion
force between an AFM tip and a hydrophilic surface: it decreases
logarithmically with the sliding velocity and vanishes at high sliding
velocities \cite{Noel12}.
Recent FFM experiments demonstrate that high humidity at low temperature
enhances the liquid lubricity while at higher temperature moisture hinders
the thermolubric effect due to the formation of liquid bridges
\cite{Gueye15}. Friction response to the dynamic lubricity in both high-
and low-temperature regimes keeps the same trends, namely the friction
force decreases with increasing the amplitude of the applied vibration on
the tip regardless of the relative humidity levels \cite{Gueye15}.

\section{Layered Materials}
\label{layered:sec}

\begin{figure}
  \centerline{\includegraphics[width=\figwid]{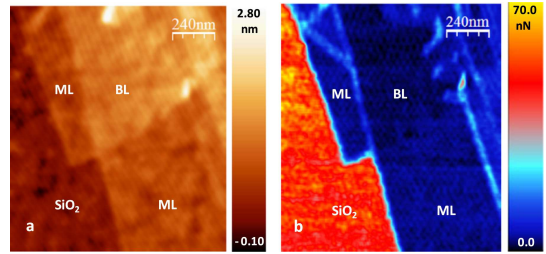}}
  \caption{\label{graphene:fig}
    (a) Contact-mode topography image obtained with a normal load of $58.8
    \pm 0.2$~nN in vacuum conditions. The region comprises a graphene
    monolayer (ML), a bi-layer (BL) and the bare SiO$_2$ substrate. (b)
    Friction force ($F_f$) map of the same region of panel (a). A marked
    difference between graphene region and the SiO$_2$ substrate is clearly
    visible and it masks the small difference between ML and BL friction
    force.
    The analysis of the friction distribution histogram shows that the
    friction signal is relatively high on SiO$_2$ ($F_f = 48 \pm 5$~nN) and
    decreases nearly ten times moving to the graphene covered regions (ML:
    $F_f = 5.4 \pm 0.7$~nN; BL: $F_f= 4.0 \pm 0.5$~nN).
    From Ref. \cite{Paolicelli15}, copyright 2015, IOP Publishing.
  }
\end{figure}

Graphene physics was awarded by Nobel Prize in 2010.  Since then,
experiments and simulations specifically devoted to tribology on graphene
and other 2D materials attracted an increasing interest. Evidence of a
large reduction of friction force (from 10 to 15 times compared to silicon
or metal-oxide surfaces, Fig.~\ref{graphene:fig}) has been accumulating
through FFM experiments, i.e. with nano-scale contacts, in a number of
different graphene systems.
These systems include graphene epitaxially grown on SiC\cite{Filliter09},
exfoliated graphene transferred on SiO$_2$ \cite{HLee09, CLee10}, suspended
graphene membranes \cite{CLee10, Deng13}, and graphene grown by chemical
vapor deposition (CVD) on metals \cite{Egberts14, Paolicelli15,
  Triphati16}.
Because of this frictional reduction, many studies indicate graphene as the
thinnest solid-state lubricant and anti-wear coating \cite{Shin11, Kim11,
  Berman14}.  Graphene is also an ideal playground for testing basic
concepts in tribology, being one of the best crystalline surfaces that is
easily accessible both experimentally and in simulation.

At the nanoscale, graphene systems exhibit a number of tribological
effects.  In few-layer systems, the importance of out-of-plane
deformations, the influence of the supporting substrate and the role of van
der Waals interactions are subject to an intense debate.
Accurate FFM measurements on few-layer graphene systems show that friction
decreases by increasing graphene thickness from a single layer up to 4-5
layers, and then it approaches graphite values \cite{Filliter09, CLee09,
  CLee10, Egberts14, Dienwiebel15}. This phenomenon, has been attributed to
the out-of-plane ``puckering '' \cite{CLee10} deformation that builds up
when the AFM tip pushes and slides over a graphene sheet. Friction
increases because graphene deformations enhance the real contact area at
the tip apex.
In this interpretation, the amount of deformation decreases with increasing
number of layers, as revealed by computer simulations \cite{LiuZhang11,
  Smolyanitsky12a, Ye12, Smolyanitsky12b}, with the result that friction
diminishes for increasing graphene thickness.

A direct evaluation of out-of-plane deformations is a difficult task
because the analysis of out-of-plane elasticity of supported 2D films
requires indentation depths smaller than the film's interlayer
spacing. Recent sub-\AA{}ngstr\"om resolution indentation measurements show
a dependence on the number of layers composing the system \cite{YGao15} and
the influence of graphene-substrate interaction as revealed originally by
FFM experiments at ``negative'' load \cite{Deng12}.
Finally, within the intense debate about proper simulation of van der Waals
interaction in h-BN and graphene, a new thickness-dependent friction
mechanism is introduced where the interlayer interactions of the sliding
top graphene sheet with the bottom layers depend on the number of layers
\cite{WGao15}.

Many efforts aim at taking advantage of the exceptional lubricant
properties of graphene for micro or macro-mechanical systems.
Millimeter-large graphene films grown on Cu and Ni by CVD and then
transferred on SiO$_2$ effectively reduce adhesion and friction forces, as
revealed by experiments performed with contact size $\sim 100~\mu$m and
loads in the tens of mN \cite{Kim11}. Wear tracks are visible at the end of
these tests and the question arises whether the pristine graphene sheet or
the accumulation of graphene debris on the counterparts are responsible for
the difference. Extensive ramping force scratch tests carried out on
exfoliated and epitaxial graphene at ambient conditions, on SiO$_2$ and SiC
respectively, exhibit a very low friction force before coating failure,
thus on presumably pristine graphene, and yield a friction coefficient of
0.03 before rupture for all graphene samples \cite{Shin11}. A similar work
on single-layer graphene grown epitaxially on SiC indicates that intact
graphene coverage provides, initially, a very low friction coefficient,
which quickly evolves to a nearly stable value five times lower than that
of SiC \cite{Marchetto12}. An analysis of the sliding track by means of
FFM, i.e.\ with nano-scale lateral resolution, reveals that the stable
lubrication regime is a mixed effect due to a covalently bound graphitic
interface layer (always present beneath graphene epitaxially grown on SiC)
and genuine graphene patches remaining attached to the substrate.

Atomistic simulations are used to shed light into the failure mechanisms
\cite{Sandoz12}. A spherical asperity with radius of 2 nm is moved over a
graphene monolayer placed in registry on a perfectly rigid
substrate. Applying a nominal load up to a few hundred nN, graphene
never delaminates and a low substrate-membrane adhesion seems to reduce
the overall damage to the graphene layer, allowing a substantial
recovery of the load-bearing capability of the graphene post
tearing. Simulations and experiments on nano indentation, sliding and
scratching of graphene covered Pt(111) surfaces are carried out using a
repulsive interaction between tip and the surface \cite{Klemenz14} and a
deformable substrate.
The simulations show, in agreement with experiments, three indentation and
sliding regimes. At low loads, the deformation is purely elastic and
sliding is almost frictionless. As the load is increased, the Pt substrate
yields plastically but graphene remains undamaged and friction increases
due to Pt plowing. Finally, graphene ruptures. The ability of graphene to
increase the load carrying capacity of this soft interface is particularly
remarkable. Additionally graphene shows a self-healing capacity during
sliding that seems to contribute to maintain stable lubrication effects.

\section{Electronic, Magnetic, Exotic, and Quantum Friction}
\label{exotic:sec}

Beside ordinary frictional mechanisms such as phonon generation,
mechanical stick-slip, viscous and viscoelastic dissipation, all the way
to wear and plastic deformation, there are also less conventional
mechanisms, connected with electronic, spin, or phase degrees of
freedom, including quantum dissipative processes.

Among them the most notable is electronic friction. Electronic friction
arises when, upon sliding on a metal surface, a tip or other moving agent
dissipates mechanical energy by exciting local currents in metals, or
electron-hole pairs in semiconductors. Since ideally the motion of a tip
alone suffices to cause electron scattering generating heat in the metal
underneath, the tip's motion will be damped when sliding on or near a metal
\cite{PerssonBook}.
In wearless nanofriction on metals, the relative importance of the phononic
and electronic contributions is still a matter of debate. To quantify this
ratio experimentally has proven difficult because the phononic and
electronic dissipation channels are generally both active.

One approach to investigate the electronic contribution to friction is to
induce large changes in the electronic density in one of the bodies in
contact. In doped semiconductors, it is possible to vary the electronic
carrier concentration in the space charge region by several orders of
magnitude by a bias voltage, essentially switching the behavior from
insulating to metallic.
With an atomic force microscope tip sliding on a silicon sample patterned
with p and n regions, a variation in friction is observed as a function of
the bias voltage \cite{Park06}: a substantial increase in friction is found
in the p-doped regions presenting a high carrier concentration near the
surface. It appears however that the main contribution to the measured
excess friction in contact is not due to the generation of electron-hole
pairs but to the force exerted by trapped charges in the oxide surface
\cite{Qi08}.

Arguably, the most direct way to estimate the relative magnitude of the
phononic and electronic contributions to friction is to work across the
superconducting transition of a metal sample.
Following private suggestions by B.N.J.\ Persson in the 1990s, this was
first experimentally attempted with a quartz-crystal microbalance technique
\cite{Dayo98}. The results show that the friction between a lead substrate
and a few-layers-thick adsorbate film made of solid nitrogen decreases by a
factor $\sim 2$ when lead becomes superconducting. However, the transition
in the friction coefficient observed in this experiment is quite abrupt, in
contrast with the predictions of the Bardeen-Cooper-Schrieffer (BCS)
theory, which estimates that thermally excited quasiparticles should lead
to a more continuous drop below Tc \cite{Persson98}.
Moreover, the same system investigated in a different QCM experiment
\cite{Renner01} with improved cryogenics and a controlled Pb surface yields
only complete pinning of the nitrogen film to the lead substrate at low
temperatures. Subsequent measurements with a more controlled set-up confirm
that nitrogen films are quite susceptible to pinning \cite{Mason01}.
Further investigations point out important differences in the frictional
response of sliding adsorbates. Experiments on nitrogen and helium films on
superconducting lead relate electronic friction to the electric
polarizability of the adsorbate species \cite{Highland06}. On the other
hand, measurements employing lighter elements such as neon, which does
slide on a lead surface even at very low temperatures \cite{Bruschi06}, do
not show any rise of electronic friction while crossing the superconducting
transition, probably because of the small polarizability of Ne atoms
\cite{Pierno10}.
More polarizable adsorbates like nitrogen, krypton and xenon are instead
found to be completely pinned to lead below 10 K \cite{Pierno11}.

\begin{figure}
  \centerline{\includegraphics[width=\figwid]{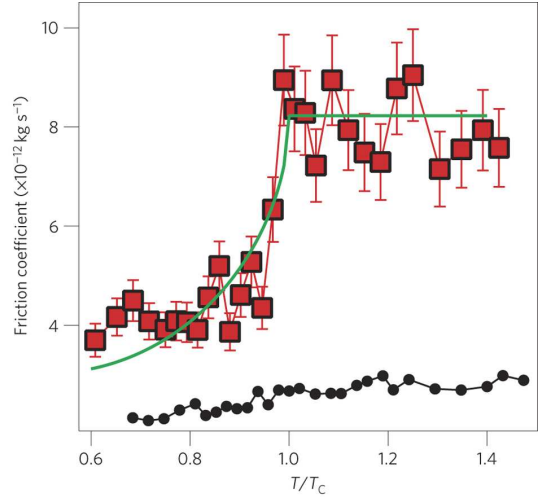}}
  \caption{\label{kisiel:fig}
    Temperature variation of the friction coefficient $\Gamma$ across the
    critical point $T_c = 9.2$~K of Nb. The red squares correspond to a
    distance $0.5$~nm between the tip and the sample. The error bars
    represent the deviation of the decay-time raw data from the exponential
    fit.
    The data are well fitted by the analytic curve expected from the BCS
    theory (green line). The black dots correspond to the temperature
    dependence of the friction coefficient $\Gamma_0$ measured at a
    separation of several micrometres (free cantilever). For figure
    clarity, the friction coefficient $\Gamma$ is shifted vertically by a
    constant ($2.5\times 10^{-12}$~kg s$^{-1}$) value. Reproduced with
    permission from Ref.~\cite{Kisiel11}. Copyright 2011, Macmillan
    Publishers Limited.
  }
\end{figure}

A convincing confirmation of electronic friction and of its suppression in
the superconducting state is provided by non-contact friction measurements
on niobium films carried out across the critical temperature using a highly
sensitive cantilever oscillating in the pendulum noncontact geometry in
ultrahigh vacuum \cite{Kisiel11}.
The friction coefficient drops by a factor between 2 and 3 when the sample
enters the superconducting state (see Fig.~\ref{kisiel:fig}). In this case,
the temperature decay of the friction coefficient is found to be in good
agreement with the BCS theory \cite{Persson98}. Noncontact friction on Nb
has an electronic nature in the metallic state, whereas phononic friction
dominates in the superconducting state. This is also supported by different
dependences of friction on the probe-sample distance and on the bias
voltage in the metallic and superconducting states.
The normal-state noncontact electronic friction may proceed in two
different ways. One is direct excitation of electron-hole pairs by the van
der Waals or Coulomb potential exerted by the tip on the metal surface
electrons. The second mechanism is the tip-induced potential-mediated
generation of a local deformation of the substrate lattice, i.e., a local
phonon, which then decays into electronic excitations of the metal, rather
than surviving and carrying energy away as in the superconducting
state. Experimental evidence does not clarify which of the two mechanisms
is at work.

The same non-contact pendulum technique is applied above the surface of
NbSe$_2$ \cite{Langer14}, a layered compound exhibiting an incommensurate
charge density wave (CDW). A multiplicity of dissipation peaks arising at
certain distances few nanometres above this surface is reported. Each peak
appears at a well-defined tip-surface interaction force of the order of
$1$~nN, and persists up to $70$~K, where the short-range order of CDWs is
known to disappear.
Comparison of the measurements with a theoretical model suggests that the
peaks are associated with local, tip-induced $2\pi$ phase slips of the CDW
incommensurate order parameter, and that dissipation maxima arise from
hysteretic behavior of the CDW phase as the tip oscillates at specific
distances where sharp local slips occur \cite{Pellegrini14}, providing an
interesting and exotic mechanism for mechanical dissipation.

Similarly, dissipation may arise by modifications of magnetic order
parameters. It is then natural to consider a possible spin-dependence of
frictional forces in the case of magnetic materials \cite{Kadau08}. To
investigate such an influence, single Co adatoms are moved over a
magnetic template by means of a spin-polarized scanning tunneling
microscopy tip \cite{Wolter12}.
It is found that the spin degree of freedom modifies the amount of
dissipated energy, the threshold force needed to move the magnetic atom and
the tip position at which the jump to the next site occurs. It may look
surprising that the spin degree of freedom can play such a significant role
in atom manipulation processes, given the different orders of magnitude of
chemical and magnetic coupling energies.
The reason is that the exchange energy does not have to
compete with the adatom binding energy, but only with the energy
barriers between adjacent adsorption sites, which can be of similar
magnitude as the exchange interaction, especially in manipulation
experiments.
Because of this similarity, magnetic adatoms can be used as local probes to
enhance the magnetic signal of atomic-scale spin structures
\cite{Ouazi14}. Single-spin magnetic dissipation phenomena are observed
with magnetic tips on, e.g., antiferromagnetic NiO \cite{Kaiser07} and are
explained theoretically as caused by quantum mechanical spin-flip events
\cite{Pellegrini10}.

There are, at least in principle, more quantum effects in nanoscale
frictional and mechanical dissipation than we can describe. Here we
mention just two. Phonon dissipation is entirely quantum at very low
temperatures but not readily simulated in that regime. Vacuum friction,
due to retardation of electromagnetic waves, and taking place because
the speed of relative motion between two sliders reduces their
generalized Casimir attraction, has also been outlined theoretically
\cite{Volokitin11, Volokitin16, Volokitin17}.

\section{Trapped Optical Systems: Ions and Colloids}
\label{optical:sec}

The fundamental understanding of the multifaceted nature of microscopic
friction, going hand in hand with the possibility of fully testing
theoretical predictions, is often hampered by both the impossibility of
tuning physical properties of real materials and the lack of
well-designed experiments at well-characterized buried interfaces.
The field of nanotribology can now benefit from the opportunities offered
by handling nano/micro particles with artificial optical potentials,
opening the possibility to change parameters almost freely and to visualize
directly the intimate mechanisms of sliding friction in simple controlled
cases.

Despite the firm theoretical background of simplified approaches such as
the Prandtl-Tomlinson and the Frenkel-Kontorova models \cite{VanossiRMP13},
describing how properties such as substrate corrugation, temperature,
driving velocity and lattice mismatch may influence the tribological
response, from intermittent stick-slip dynamics to superlubric regimes of
motion, neither of these models has been directly tested experimentally.

Thanks to a highly innovative experimental apparatus \cite{Bohlein12}, 
a brand new light is cast on elemental tribological processes by exploiting 
the versatility of charged colloidal systems driven across interfering 
laser-generated potentials, whose geometry can be tuned at will.
While AFM, SFA and QCM provide, a system response in terms of crucial, but
averaged, physical quantities, colloidal friction provides an unprecedented
real-time insight into the basic dynamical mechanisms at play, excitingly
observing what each individual particle is directly doing at the sliding
interface.

\begin{figure}
  \centerline{\includegraphics[width=\figwid]{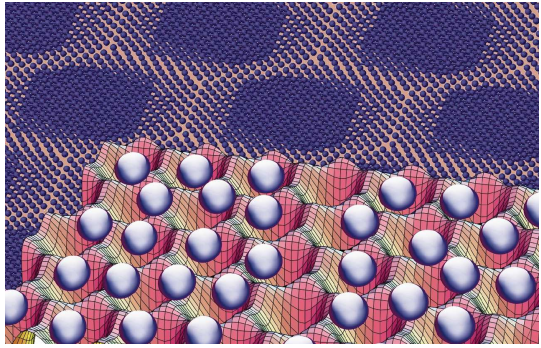}}
  \caption{\label{colloid:fig}
    Zoom-in (front) of a MD simulated frictional interface between a
    colloid monolayer and an optical corrugated substrate potential. The
    overlayer/substrate lattice mismatch, tunable experimental parameter,
    realizes network of solitonic structures (back), ruling the system
    tribological response.
  }
\end{figure}

Specifically, by driving highly charged polystyrene spheres,
naturally forming in water a 2D triangular crystal \cite{Baumgartl04, Baumgartl07}, 
across both a commensurate and incommensurate laser-generated substrate geometry, 
the colloid approach \cite{Bohlein12} highlights the crucial tribological role played by
localized superstructures (such as kinks and antikinks), which emerge as
shadow-like density modulations in crystalline overlayers that are out of
registry with their substrates.
According to theory and numerical simulations \cite{Vanossi12PNAS, Hasnain14, Vanossi15, 
Paronuzzi16}, the experiment shows the dramatic change of the static-friction threshold 
from a strongly pinned colloidal regime to an almost superlubric frictional sliding as a function of the
overlayer/substrate lattice mismatch (see Fig.~\ref{colloid:fig}).

While nucleation dynamics rules the depinning mechanism of a stiff
commensurate colloidal monolayer \cite{Hasnain14}, for an incommensurate
interface the presence/absence of pinning depends upon the system
parameters; when an increasing substrate corrugation turns an initially
free-sliding network of solitons into a colloid pinned state, the static
friction force crosses a well-defined, Aubry-like, dynamical phase
transition, from zero to finite \cite{Vanossi12, MandelliPRB15}. The transition
value for the critical corrugation depends significantly upon the
relative colloid/substrate orientation, which, energetically, is always
slightly misaligned, as shown in recent work \cite{MandelliPRL15}. By further
considering an optical substrate with quasiperiodic symmetry
\cite{Bohlein12PRL}, which lacks translational invariance, the colloidal
approach shows how the peculiar phenomenon of directional locking may
also occur on overlayers driven on quasicrystalline potentials
\cite{Reichhardt11}.

\begin{figure}
  \centerline{\includegraphics[width=\figwid]{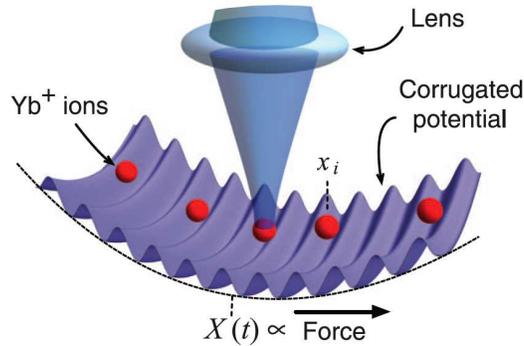}}
  \caption{\label{iontrap:fig}
    A sketch of the synthetic nanofriction interface between a Coulomb
    crystal of $^{174}$Yb$^+$ ions and an optical corrugated lattice, with
    imaging realized through a microscope with single-ion resolution.
    From Ref. \cite{Bylinskii15}, copyright 2015, The American Association
    for the Advancement of Science.
  }
\end{figure}

From colloidal mesoscale suspensions down to the nanoworld of cold ions,
the use of artificial tribology emulators has recently taken us to the
ultimate limit of atomic friction. Inspired by earlier theoretical
suggestions \cite{Benassi11, GarciaMata07, Mandelli13}, and as predicted by
many-particle models \cite{VanossiRMP13, Manini15, Manini16}, the
experimental setup of a laser-cooled Coulomb crystal of ions moving over a
periodic light-field potential highlights the practical feasibility to
control friction, from strongly dissipative stick-slip to almost free
sliding, by tuning the interface structural mismatch \cite{Bylinskii15} and
the optical corrugation \cite{Bylinskii16} at the level of just a few
interacting atom system (see Fig.~\ref{iontrap:fig}).

Compared to standard experimental tribology techniques, another
outstanding achievement within the framework of such ion-crystal system
in a optical lattice relies on the possibility to span almost five
orders of magnitude in velocity, while controlling temperature and
dissipation \cite{Gangloff15}, emulating the PT model to near perfection.

\section{Conclusions}
\label{conclusions:sec}

The physics of tribology from the atomic, to the nano and micro scale is
alive and well.
In this bird's eye review there is of course neither space nor scope for
dozens of exciting ongoing research lines.
Yet, we hope to have given at least the flavor of some new and interesting
problems that are being discovered and studied, with the interplay of
experiments, theory and simulations that are making this area currently hot
and rich.
Last but not least, the efforts that are being made to develop the physical
understanding will undoubtedly contribute in due course to novel 
methodologies in applied tribology.

\section*{Acknowledgments}

Discussion and collaboration with
A.\ Benassi, P.P.\ Baruselli, A.R.\ Bishop, O.M.\ Braun, O.\ Brovko,
R.\ Capozza, Y.\ Crespo, L.\ Gigli, R.\ Guerra, A.\ Laio, F.P.\ Landes,
D.\ Mandelli, E.\ Panizon, F.\ Pellegrini, B.N.J.\ Persson, M.\ Peyrard,
S.\ Prestipino, G.E.\ Santoro, J.\ Scheibert, M.\ Teruzzi, M.\ Urbakh,
T.\ Zanca, S.\ Zapperi is gratefully acknowledged.
This work is partly funded by
the ERC Advanced Grant No. 320796-MODPHYSFRICT,
and by COST Action MP1303. G.P.  also acknowledges 
Regione Emilia Romagna, Project INTERMECH ? MO.RE.

\bibliographystyle{advphysx}

\end{document}